\documentclass[12pt,a4paper]{article}
\usepackage{amsmath, amssymb, fullpage, graphics}
\usepackage{caption, graphicx, fancyhdr, fancybox}
\usepackage[cp866]{inputenc}
\usepackage{color}

\newcommand{\bv}{\mathbf{v}}
\newcommand{\bx}{\mathbf{x}}
\newcommand{\bg}{\mathbf{g}}
\newcommand{\bA}{\mathbf{A}}
\newcommand{\bU}{\mathbf{U}}
\newcommand{\bF}{\mathbf{F}}

\newcommand{\br}{\mathbf{r}}
\newcommand{\bq}{\mathbf{q}}

\begin{document}

\begin{center}
{\sf STABILITY ANALYSIS OF SHEAR FLOWS IN A HELE-SHAW CELL} \\
\vspace{2mm}

A.\,A. Chesnokov$^{1,2}$, I.\,V. Stepanova$^{3}$ \\[2mm]

${^1}$Novosibirsk State University, Novosibirsk, Russia \\
Pirogova Str. 2, Novosibirsk, 630090, Russia \\[2mm]
${^2}$Lavrentyev Institute of Hydrodynamics SB RAS, \\
Lavrentyev Ave. 15, Novosibirsk, 630090, Russia \\[2mm] 
$^3$Institute of Computational Modeling SB RAS, \\ 
Krasnoyarsk, Akademgorodok, 660036, Russia \\[2mm]

e-mails: chesnokov@hydro.nsc.ru, stepiv@icm.krasn.ru
\end{center}

\begin{abstract}
A mathematical model describing motion of an inhomogeneous incompressible fluid in a Hele-Shaw cell is considered. Linear stability analysis of shear flow class is provided. The role of inertia, linear friction and impermeable boundaries in Kelvin--Helmholtz instability development is studied. Hierarchy of simplified one-dimensional models of viscosity- and density-stratified flows is obtained in long-wave approximation. Interpretation of Saffman--Taylor instability development is given in the framework of these models.

\end{abstract}


\section{Introduction} 

A general property of fluid shear flows is the Kelvin--Helmholtz instability. This type of the interface instability of ideal fluid layers moving with different velocities is studied clearly enough. Classical examples of linear analysis of such flows are described in \cite{Dr05, Dikii} and in others monographs. Recent review \cite{Sahu} is devoted to stability analysis of viscosity-stratified shear flows. A substantial
theoretical object having practical applications is the Hele-Shaw shear flows \cite{Gondret}. Displacing more viscous fluid by less viscous one leads to formation of viscous fingers. It is caused by Suffman--Taylor instability~\cite{Saffman}. Classical models of fluid motions in a porous medium are based on linear Darcy law \cite{TanHomsy, Homsy}. However shear instability can develop on the longitudinal interface in the case of fast flow at formed viscous fingers also \cite{Zvyagin}. It is necessary to use the nonlinear Darcy law for analysis of this phenomenon. It takes into account inertia terms in the momentum equations. Such models are considered in \cite{Chevalier, DiasMiranda}. There are exact and experimental studies of viscous fingers formation with inertia effect for immiscible fluids in the papers cited. The same problem for miscible fluids is described in \cite{YuanAzaiez}. It is shown that inertia slows down the growth of viscous fingers. Effect of Hele-Shaw cell constriction and walls elasticity on Saffman--Taylor instability is analyzed in \cite{Housseiny, Pihler}.

In the manuscript the nonlinear system of equations describing inhomogeneous fluid flow in a Hele-Shaw cell is under study. The model admits a class of shear flows. The class is characterized by arbitrary dependence of the horizontal velocity component and density on the vertical spatial coordinate. One of the main purpose of the study is stability analysis of this class of flow. The equation for amplitudes of small perturbations is obtained. The conditions on the interface are formulated for fluids with different physical properties. Stability analysis of two-layer Hele-Shaw flow is carried out in details. It is shown that the interface of layers moving with different velocities is instable for short-wave perturbations and stable for long-wave ones.

The pressure changes weakly in the vertical direction for prevailing horizontal fluid motion  \cite{ChesnLiap}. It allows to consider long-wave approximation and class of layered flows. A hierarchy of one-dimension models is constructed basing on two simplifications of the momentum equations. They are linearization and using of Darcy law. Two-layer flow of viscosity- and density-stratified fluid is studied. Additionally the evolutionary form of the governing equations is obtained both in multi-layer and two-layer cases. The hyperbolicity property is proved for the models above. It enables to apply Godunov method modifications for numerical calculations. In particular Nessyahu--Tadmor central scheme is used in the paper. Instability of the interface is proved at displacing more viscous fluid by less viscous one if density stratification is neglected.

The paper is organized as follows. In Section 2 we introduce the mathematical model of the process described above. In Section 3 stability analysis of shear flows is carried out. The cases of homogeneous and stratified flows are studied. In Section 4 we consider layered flows in long-wave approximation. We describe two simplified models of such flows which are treated both analytically and numerically.

\section{Mathematical model}

Equations of motion of a viscous incompressible fluid in a Hele-Shaw cell (region between two parallel plates separated by a small gap) have the form
\begin{equation}\label{eq:N-St}
 \begin{array}{l}\displaystyle
   \rho (\bv_t+(\bv\cdot\nabla)\bv)+\nabla p= \mu\bv_{zz}+\rho\bg, \quad \nabla\cdot\bv=0.
  \end{array}
\end{equation}
Here $\bx=(x,y,z)$ are the coordinate vector, $t$ is the time, $\bv=(u,v,w)$ is the velocity  vector, $p$ is the pressure, $\rho$ is the density, $\mu$ is the viscosity and $\bg~=~(0,-g,0)$ is the acceleration of gravity vector. The operator $\nabla$ is calculated with respect to the coordinate vector. The characteristic sizes of cell $(L,H)$ in the $x$ and $y$ direction respectively are significantly higher than the cell gap $b$. That is why the summands $\bv_{xx}$ and $\bv_{yy}$ vanish in the momentum equations. They are negligible compared to $\bv_{zz}$. We assume the density and viscosity are monotonic dependencies of concentrations $\varkappa$: $\mu=\mu(\varkappa)$ and $\rho=\rho(\varkappa)$. The variable $\varkappa$ varies from zero to one and satisfies the equation
\begin{equation}\label{eq:trajectory}
  \varkappa_t+\mathbf{v}\cdot\nabla\varkappa=0.
\end{equation}
It should be noted that $\varkappa$ can change continuously or take piecewise constant values. The first case corresponds to an inhomogeneous fluid, the second one characterizes a multicomponent fluid.

We consider the velocity field in the form
\[ u=\frac{3}{2}\Big(1-\Big(\frac{2z}{b}\Big)^2\Big)u'(t,x,y), \quad
   v=\frac{3}{2}\Big(1-\Big(\frac{2z}{b}\Big)^2\Big)v'(t,x,y), \quad w=0. \]
It provides the fulfillment of no-slip conditions on the cell walls $z=\pm b/2$. We also suppose that the functions $p$ and $\varkappa$ do not depend on $z$. Integrating equations (\ref {eq:N-St}) and (\ref{eq:trajectory}) from $-b/2$ to $b/2$ leads to the system
\begin{equation}\label{eq:model}
 \begin{array}{l}\displaystyle
  \rho(u_t+\beta(uu_x+vu_y))+p_x=-\mu u, \\[3mm]\displaystyle
  \rho(v_t+\beta(uv_x+vv_y))+p_y=-\mu v-\rho g, \\[3mm]\displaystyle
  u_x+v_y=0, \quad \varkappa_t+u\varkappa_x+v\varkappa_y=0.
 \end{array}
\end{equation}
The primes are omitted. Here and below $\mu$ denotes the modified fluid viscosity $12\mu/b^2$. The coefficient $\beta$ is equal to 6/5.  The simplifications above are connected with inequalities $b<<L$, $b<<H$ and often used for modeling of Hele-Shaw flows. For the constant density and using of linear Darcy law equations (\ref{eq:model}) reduce to the classical problem without diffusion consideration~\cite{Homsy}. In the framework of the model described the interface instability is provided if displaced fluid is more viscous than the displacing one.

The following Sections are devoted to stability analysis of steady shear solutions of equations (\ref{eq:model}). Also a class of layered flows is considered in long-wave approximation.

\section{Stability analysis of shear flows}

Equations (\ref{eq:model}) admit the class of solutions
\begin{equation}\label{eq:base-sol}
  u=U(y), \quad v=0, \quad  p=P(y)-\alpha x, \quad \varkappa=C(y),
\end{equation}
where $\alpha>0$ is constant, $U$, $P$ and $C$ are arbitrary smooth functions. The following relations are valid
\[ \mu=\mu(C)=\frac{\alpha}{U(y)}, \quad \rho=\rho(C)=R(y), \quad P'(y)=-gR(y). \]
The vector of velocity components, pressure and concentration of perturbed flow deviate from the basis flow weakly. That is why these values take the form
\begin{equation}\label{eq:perturbate-sol}
  u=U(y)+\tilde{u},\quad v=\tilde{v},\quad p=P(y)-\alpha x+\tilde{p},
  \quad \varkappa=C(y)+\tilde{\varkappa}.
\end{equation}
Here $\tilde{u}$, $\tilde{v}$, $\tilde{p}$ and $\tilde{\varkappa}$
are the functions of all independent variables. Dependencies $\mu=\mu(\varkappa)$ and $\rho=\rho(\varkappa)$ have the form
\[ \mu=\frac{\alpha}{U(y)}\Big(1-\frac{U'(y)\tilde{\varkappa}}{U(y)C'(y)}\Big), \quad
   \rho=R(y)+\frac{R'(y)\tilde{\varkappa}}{C'(y)}\,. \]
Substitution of solution (\ref{eq:perturbate-sol}) into equations (\ref{eq:model}) and linearization of the system obtained give the equations for small perturbations finding
\begin{equation}\label{eq:mod-linear}
 \begin{array}{l}\displaystyle
  u_t+\beta(Uu_x+U'v)+\frac{1}{R}p_x=\frac{\alpha}{UR}\Big(\frac{U'\varkappa}{C'}-u\Big), \\[4mm]\displaystyle
  v_t+\beta Uv_x+\frac{1}{R}p_y=-\frac{1}{R}\Big(\frac{\alpha v}{U}+\frac{gR'\varkappa}{C'}\Big), \\[4mm]\displaystyle u_x+v_y=0, \quad \varkappa_t+U\varkappa_x+vC'=0.
 \end{array}
\end{equation}
The tildes are omitted. The prime denotes the derivative with respect to $y$.

We use the stream function $\psi(t,x,y)$ for the perturbations of the basic flow. It satisfies the equalities $u=\psi_y$ and $v=-\psi_x$. We differentiate the first equation in (\ref{eq:mod-linear}) with respect to $y$ and the second one with respect to $x$. Then we eliminate the pressure $p$ and obtain the system for perturbations of the stream function and the concentration only
\begin{equation}\label{eq:psi-rho-mu}
 \begin{array}{l}\displaystyle
  (\partial_t+\beta U\partial_x)\Big(\Delta\psi+\frac{R'}{R}\psi_y\Big)-
  \Big(U''+\frac{R'U'}{R}\Big)\beta \psi_x=\frac{gR'}{RC'}\varkappa_x- \\[5mm]\displaystyle
  \quad\quad -
  \frac{\alpha U'}{RU^2} \Big(\frac{U'\varkappa}{C'}-\psi_y\Big)+\frac{\alpha}{UR}\bigg(\Big(U''-\frac{U'C''}{C'}\Big)\frac{\varkappa}{C'}
  +\frac{U'}{C'}\varkappa_y-\Delta\psi\bigg)\,, \\[5mm]\displaystyle
  (\partial_t+U\partial_x)\varkappa-C'\psi_x=0.
 \end{array}
\end{equation}
A solution of these equations is sought in the form
\begin{equation}\label{eq:wave-sol}
  (\psi, \varkappa)=(\Psi(y), M(y))\exp{(ik(x-ct))},
\end{equation}
where $c$ is the phase velocity, $k>0$ is the wave number, $2\pi/k$ is the wave length and $i$ is the imaginary unit. Substitution of formulas (\ref{eq:wave-sol}) into (\ref{eq:psi-rho-mu}) leads to equations for perturbation amplitudes $\Psi(y)$ and $M(y)$. Due to the second equation in (\ref{eq:psi-rho-mu}) we have $M=(U-c)^{-1}C'\Psi$. The first equation in (\ref{eq:psi-rho-mu}) reduces to the second order ODE for the function $\Psi(y)$
\begin{equation}\label{eq:psi-main}
 \begin{array}{l}\displaystyle
  \Big(\beta U-c-\frac{i\alpha}{kUR}\Big)\big(\Psi''-k^2\Psi\big)-
  \Big((U-c)\beta-\frac{i\alpha}{kUR}\Big)\frac{U''\Psi}{U-c}= \\[5mm]\displaystyle
  \hspace{-3mm}=-\frac{i\alpha
  U'}{kRU^2}\,\frac{2U-c}{U-c}\Big(\Psi'-\frac{U'\Psi}{U-c}\Big)\!+\!
  \bigg(\Big(\frac{g}{U-c}\!+\!\beta U'\Big)\Psi\!-\!(\beta U-c)\Psi'\bigg)\frac{R'}{R}.
 \end{array}
\end{equation}
If the flow region is restricted by the horizontal walls $y=\pm h$ then the boundary conditions for equation (\ref{eq:psi-main}) have the form $\Psi(-h)=\Psi(h)=0$ and correspond to impermeability conditions on the rigid walls. In the absence of walls the limit relations $\Psi\to 0$ fulfill at $y\to\pm\infty$.

\subsection{Conditions on the interface}

We consider shear flow (\ref{eq:base-sol}) with discontinuity at $y=0$. We do not take into account the surface tension. The limit values $U^\pm$ and $R^\pm$ of functions $U(y)$ and $R(y)$ are
different at $y\to \pm 0$. The superscript ``+'' corresponds to the upper fluid. The linearized kinematic condition takes the form
\[ (\partial_t+U^\pm\partial_x)\zeta=v^\pm=-\psi_x^\pm, \]
where $\zeta(t,x)$ is the small perturbation of the interface. Eliminating the function $\zeta(t,x)$ from the latter equation gives the formula $(\partial_t+U^+\partial_x)\psi_x^-=(\partial_t+U^-\partial_x)\psi_x^+$.
For wave solutions (\ref{eq:wave-sol}) this formula is rewritten as
\begin{equation}\label{eq:kin-bc}
  (U^+-c)\Psi^-=(U^--c)\Psi^+.
\end{equation}

On the interface the linearized dynamic condition has the form
\[ [p-gR\zeta]^+_-=0, \]
where $[f]^+_-=f^+-f_-$ is the difference between limit values at the discontinuous. In order to obtain the dynamic condition for the stream function we apply the differential operator $(\partial_t+U^+\partial_x)\partial_x$ to the latter equation. Taking into account equations (\ref{eq:mod-linear}) the kinematic condition on the interface and form of solution (\ref{eq:wave-sol})
we have the formula
\begin{equation}\label{eq:din-bc}
  \bigg[\Big((\beta U-c)R-\frac{i\alpha}{kU}\Big)\Psi'-
  \Big((U-c)\beta
  R-\frac{i\alpha}{kU}\Big)\frac{U'\Psi}{U-c}\bigg]^+_-=\frac{g\Psi^+}{U^+-c}\big[R\big]^+_-.
\end{equation}

The solution of equation (\ref{eq:psi-main}) and conditions (\ref{eq:kin-bc}), (\ref{eq:din-bc}) allow to obtain the dispersion relation $c=c(k)$. Analysis of this relation gives the conclusion
about the flow stability. It should be noted that at $\alpha=0$, $\beta=1$ formulas (\ref{eq:psi-main}), (\ref{eq:kin-bc}) and (\ref{eq:din-bc}) reduce to the classical relations of linear stability theory of plane-parallel shear flows of ideal fluid \cite{Dr05, Dikii}.

\subsection{Stability analysis of two-layer flow}

\begin{figure}[h]
\begin{center}
\resizebox{1\textwidth}{!}{\includegraphics{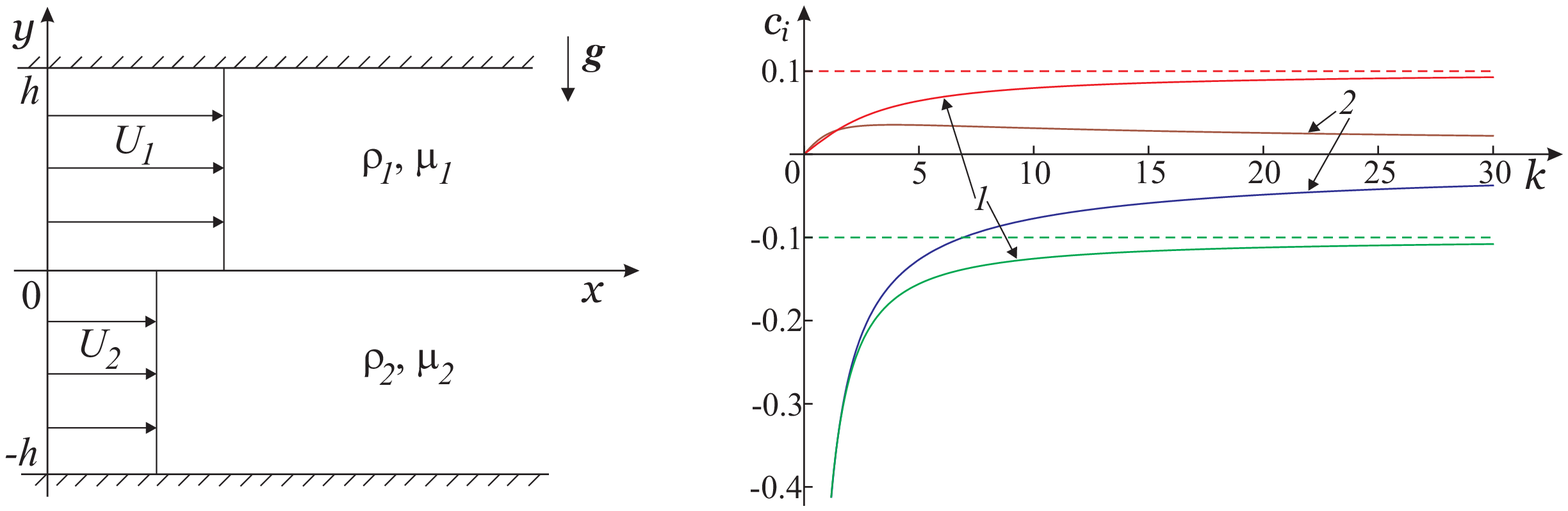}}\\[2mm]
\parbox{0.46\textwidth}{\caption{{\small Two-layers flow scheme.}} \label{fig:1}} \hfill
\parbox{0.50\textwidth}{\caption{{\small Dispersion curves ($U_1/U_2\!=\!6/5$, $\alpha\!=\!1/2$):
 {\it 1}~---~$\beta=1$; {\it 2}~---~$\beta=6/5$.}} \label{fig:2}}
\end{center}
\end{figure}

We consider two-layer motion of fluid with the interface $y=0$ between impermeable walls $y=\pm h$. The flow scheme is presented in Fig.~\ref{fig:1}. The fluid layers move with constant positive
velocities $U_1$, $U_2$, have constant viscosities $\mu_j$ and densities $\rho_j$ ($j=1,2$). The compatibility condition $U_j\mu_j=\alpha$ follows from formulas (\ref{eq:base-sol}) and is
assumed to be valid. In the case of piecewise constant distribution of velocity and density equation (\ref{eq:psi-main}) takes the form $\Psi''-k^2\Psi=0$. We solve this equation with conditions $\Psi(\pm h)=0$ and use formulas (\ref{eq:kin-bc}), (\ref{eq:din-bc}). As the result we have the dispersion relation
\begin{equation}\label{eq:disp-rel}
 \begin{array}{l}\displaystyle
   \big((\beta U_1-c)k\rho_1-i\alpha U_1^{-1}\big)(U_1-c)+
   \big((\beta U_2-c)k\rho_2- \\[2mm]\displaystyle
  \quad\quad\quad\quad\quad -i\alpha U_2^{-1}\big)(U_2-c)-g(\rho_2-\rho_1)\varphi(h,k)=0,
 \end{array}
\end{equation}
where $\varphi(h,k)=\tanh(2hk)$. When the impermeable walls are absent, i.e. $\Psi(y)\to 0$ at $y\to\pm\infty$, the dispersion relation has also form (\ref{eq:disp-rel}) but $\varphi(h,k)=1$.
Formula (\ref{eq:disp-rel}) connects the complex phase velocity $c=c_r+ic_i$ and wave number $k$ at given values of $U_j$, $\rho_j$, $\alpha$ and $\beta$. If there are such values of parameter $k$ for
which equation (\ref{eq:disp-rel}) has complex roots with positive imaginary part $c_i$ then the considered flow is instable.

\subsection{Two-layer flow of a homogeneous fluid}

We consider two-layer fluid with identical densities $\rho_1=\rho_2=1$. Then the gravity force and walls do not influence on the stability of the flow studied. Following \cite{Gondret} the coefficient $\beta=6/5$ is often considered to be equal to one. When $\alpha=0$ ($\alpha=\mu_1 U_1=\mu_2 U_2$) and $\beta=1$ we have $c_i={\rm Im}\,c=\pm(U_1-U_2)/2$ from dispersion relation (\ref{eq:disp-rel}). These dependencies are presented by dashed lines in Fig.~\ref{fig:2} and correspond to instability of the contact discontinuity in homogeneous ideal fluid. When $\alpha>0$
the roots of quadratic equation (\ref{eq:disp-rel}) depend on the wave number $k$. Special dependencies $c_i=c_i(k)$ are presented in Fig.~\ref{fig:2} (lines {\it 1}). The plots are obtained at parameters $U_1=6/5$, $U_2=1$, $\alpha=1/2$ and $\mu_j=\alpha/U_j$. The values are pointed out for the CGS system. Also they can be assumed to be dimensionless. It is justified because we are
interested in qualitative behavior of solutions. The upper branch of the graph $c_i=c_i(k)$ increases monotonically from zero to $(U_1-U_2)/2$. The long waves ($k\to 0$) are stable neutrally while perturbations of short waves grow with the same velocity as at $\alpha=0$.

When $\beta=6/5$ the dispersion curves $c_i=c_i(k)$ have the same qualitative behavior as at $\beta=1$ (lines {\it 1} in Fig.~\ref{fig:2}) if the following inequality fulfills
\[ f\equiv(1+\beta)^2(U_1+U_2)^2-8\beta(U_1^2+U_2^2)<0. \]
At $\beta=1$ this inequality holds because $-(U_1-U_2)^2<0$. One can readily see that at $\beta>1$ and considerably small difference $|U_1-U_2|$ the inverse inequality $f>0$ is valid. The roots of equation (\ref{eq:disp-rel}) take the form
\[ c_{1,2}=\frac{1}{4}\bigg((1+\beta)(U_1+U_2)\pm\sqrt{f}\bigg) \]
in the case of $\alpha=0$. It corresponds to instability of two-layer ideal flow with correction factor $\beta>1$ in the momentum equations. At $\alpha>0$, $\beta=6/5$ and $f>0$ the inherent
velocity behavior is presented in Fig.~\ref{fig:2} (lines {\it 2}). In this case the greatest growth of disturbances occurs in the middle range of wavelengths.

It is interesting that at $\beta=0$ the flow is stable. Applying Vieta theorem to the quadratic equation
\[ c^2-\frac{U_1+U_2}{2}\Big(1-\frac{i\alpha}{kU_1U_2}\Big)c-\frac{i\alpha}{k}=0 \]
proves that there are no complex roots with positive imaginary part. Actually let $c_1=c_{1r}+ic_{1i}$ and $c_2=c_{2r}+ic_{2i}$ be the roots of the latter equation. Then from Vieta theorem we conclude the fulfillment of the following inequalities
\[ c_{1i}/c_{2r}=c_{1r}/c_{2i}<0, \quad c_{1r}/c_{1i}=c_{2i}/c_{2r}<0. \]
Due to the inequalities $c_{1r}+c_{2r}=(U_1+U_2)/2>0$, $c_{1r}>0$ and/or $c_{2r}>0$ we infer that $c_{1i}<0$ and $c_{2i}<0$.

\subsection{Two-layer flow of a stratified fluid}

\begin{figure}[h]
\begin{center}
\resizebox{1\textwidth}{!}{\includegraphics{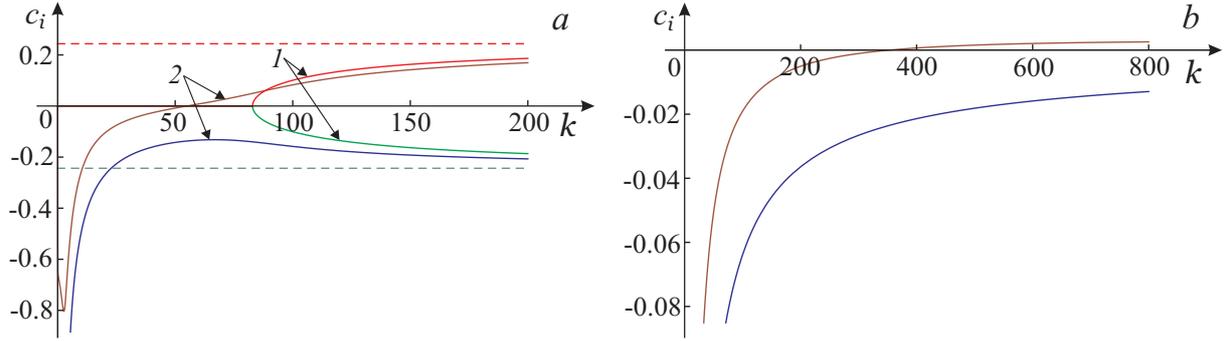}}\\[2mm]
{\caption{\small{Dispersion curves: {\it a} --- $\alpha=0$ (curves {\it 1}) and $\alpha=9$ (curves {\it 2})  at $U_1/U_2=3/2$; {\it b} --- $\alpha=9$ at $U_1/U_2=6/5$.}} \label{fig:3}}
\end{center}
\end{figure}

We consider the shear flow (Fig.~\ref{fig:1}) with density stratification $\rho_2=1$, $\lambda=\rho_1/\rho_2<1$. The function $\varphi(h,k)$ varies in the interval $(0,1)$ and makes adjustments to the last summand of the left part in (\ref{eq:disp-rel}). It does not influence on the qualitative results. At $\alpha=0$ dispersion relation (\ref{eq:disp-rel}) has the following roots
\[ \begin{array}{l}\displaystyle
    c=\frac{1}{1+\lambda}\Big(\frac{1+\beta}{2}(\lambda U_1+U_2)\pm
    \\[3mm]\displaystyle
    \quad\quad\quad\quad
    \pm\sqrt{\frac{(1+\beta)^2}{4}(\lambda U_1+U_2)^2-(\lambda+1)(\lambda U_1^2+U_2^2)\beta+
    \frac{g\varphi}{k}(1-\lambda^2)}\Big).
    \end{array} \]
As in the previous example there is difference between the cases $\beta=1$ and $\beta=6/5$. At $\beta=1$ and $\alpha=0$ the flow is instable always because at $k\to\infty$ the following relation is valid
\[ {\rm Im}\,c\to \mp (U_1-U_2)(1+\lambda)^{-1}\sqrt{\lambda}. \]
It is pointed out in Fig.~\ref{fig:3},{\it a} by dashed lines. At $\beta=6/5$ there are parameters for stable flow.

Fig.~\ref{fig:3},{\it a} presents the dispersion curves obtained from solution of equation (\ref{eq:disp-rel}) for the parameters $\beta=1.2$, $U_1=1.5$, $U_2=1$, $g=980$, $h=10$ and
$\lambda=0.99$. Lines~{\it 1} correspond to $\alpha=0$ and lines~{\it 2} are pictured for $\alpha=9$. The stability area is essentially wider compared to the homogeneous flow because of the
positive values $c_i$ only for $k>k_*$, where $k_*$ is the critical value. In the example under study $k_*\approx 83$ at $\alpha=0$ and $k_*\approx 55$ for $\alpha\neq0$. The values $k_*$ increase
monotonically with the growth of the shift value $|U_1-U_2|$. When the parameter $\alpha$ decreases curves~{\it 2} draw close to lines~{\it 1}. We note that at $\beta=1$ the qualitative behavior of
dispersion curves is the same. The values of critical wave numbers coincide for $\alpha=0$ and $\alpha\neq0$. For the example considered these values are $k_*\approx 79$ and $76$. We take
$U_1=6/5$ and the other parameters do not change. This case corresponds to stability of flow for $\alpha=0$ due to the positive radicand in formula for $c(k)$ for all $k>0$. The flow with $\alpha\neq0$ loses stability for the sufficiently large values $k$. In the case considered $k_*\approx 353$.

After analysis described above we conclude that long-wave perturbations ($k\to 0$) are stable in a viscous fluid ($\alpha>0$). It gives grounds for the study of layered flows in long-wave approximation in details.

\section {Layered flows}

We consider the motion of an incompressible fluid in a Hele-Shaw cell. Its horizontal size
 $L$ is much more than its depth $H$, i.e. $H/L=\varepsilon<<1$. Further we assume
\[ t\to \varepsilon^{-1} t, \quad x\to \varepsilon^{-1} x,
   \quad  v\to \varepsilon v, \quad \mu\to \varepsilon \mu \]
and omit all terms of order $\varepsilon^2$ in equations (\ref{eq:model}). It corresponds to the long-wave approximation \cite{LT00, ChL11}. As the result we have the model
\begin{equation}\label{eq:long_wave_app}
  \begin{array}{l}
  \rho(u_t+\beta(uu_x+vu_y))+p_x=-\mu u, \quad p_y=-g\rho, \\[2mm]\displaystyle
  u_x+v_y=0, \quad \varkappa_t+u\varkappa_x+v\varkappa_y=0; \\[2mm]\displaystyle
  \rho=\rho(\varkappa), \quad \mu=\mu(\varkappa); \quad v\big|_{y=0}=v\big|_{y=H}=0.
 \end{array}
\end{equation}
Here $y=0$ and $y=H$ are the lower and upper boundaries in the $y$-direction. Note that stability of Hele-Shaw shear flows for weakly compressible barotropic fluid for $\beta=1$ is studied in \cite{MedChesn} in the framework of hyperbolicity of integro-differential motion equations \cite{LT00, ChL11}. However for the long-wave models of incompressible inhomogeneous fluid the transition to the semi-Lagrangian variables does not lead to simplifications of motion equations and does not apply here.

We deal with the class of layered flows
\[ u=u_i(t,x), \quad \varkappa=\varkappa_i=\mathrm{const}, \quad
   \big(y\in (y_{i-1},y_i), \quad i=1,\ldots,N\big), \]
wherein $0=y_0<y_1(t,x)<\ldots<y_N(t,x)=H$. For this class equations (\ref{eq:long_wave_app}) take the form
\begin{equation}\label{eq:layers}
\begin{array}{l}
    \rho_i\Big(u_{it}+\beta u_i u_{ix}+g\sum\limits_{j=1}^i h_{jx}\Big)+
    g\sum\limits_{j=i+1}^N h_{jx}=-p_{0x}-\mu_i u_i, \\[4mm]\displaystyle
    h_{it}+(u_i h_i)_x=0, \quad \sum\limits_{i=1}^N h_i=H, \quad \sum\limits_{i=1}^N u_i
    h_i=Q,
   \end{array}
\end{equation}
where $h_i=y_i-y_{i-1}$ is the depth of layer $i$ with the density $\rho_i$ and the velocity $u_i$; $Q$ is the constant flow rate through the vertical section of the cell; $p_0$ is the pressure on
the upper lid. For the derivation of the equations we use the kinematic condition on the interface. The similar models for unstratified fluid ($\rho_i=\rho={\rm const}$) are studied in \cite{ChesnLiap}.

System (\ref{eq:layers}) is reduced to evolutionary form using new dependent variables $s_i=\rho_i u_i-\rho_N u_N$. Then the motion equations are rewritten as follows
\[\begin{array}{l}\displaystyle
    s_{it}+\frac{\beta}{2} \Bigg(\rho_iu_i^2+\rho_Nu_N^2+2g\rho_i
    \sum\limits_{j=1}^i h_j-2g\rho_N\sum\limits_{j=1}^N h_j+
    2g\sum\limits_{j=i+1}^{N-1} h_j\Bigg)_x=\\[6mm]\displaystyle
    \hspace{9cm}=-\mu_i u_i+\mu_N u_N, \\[2mm]\displaystyle
    h_{it}+(u_i h_i)_x=0, \quad \sum\limits_{i=1}^N h_i=H, \quad \sum\limits_{i=1}^N u_i h_i=Q,
    \quad i=1,\ldots,N-1.
    \end{array}\]
Depth and velocity of the last layer have the form
\[ h_N=H-\sum\limits_{j=1}^{N-1}h_j,\quad
   u_N=\Big(H+\sum\limits_{j=1}^{N-1} \Big(\frac{\rho_N}{\rho_j}-1\Big)h_j\Big)^{-1} \Big(Q-\sum\limits_{j=1}^{N-1}\frac{s_jh_j}{\rho_j}\Big). \]

In some cases it is suitable to use a moving coordinate system. It has the average flow velocity $U=Q/H$. The corresponding change of variables is $x'=x-Ut$, \ $u'=u-U$, \ $U={\rm const}$. Then the
first equation in (\ref{eq:layers}) takes the form
\[ u_{it}+\big(\beta u_i+
   (\beta-1)U\big)u_{ix} +\frac{g}{\rho_i}\sum\limits_{j=1}^{i}h_{jx}
   +\frac{1}{\rho_i}p_{0x}=-\frac{\mu_i}{\rho_i}\big(u_i+U\big). \]
The primes are omitted. The other equations in (\ref{eq:layers}) do not vary. Further we consider modifications of system (\ref{eq:layers}) and make numerical calculations of the interface position between two fluids with different physical properties. In particular it is discussed that interpretation of Saffman--Taylor instability can be given with the help of the one-dimension models.

\subsection{Two-layer flow of stratified fluid}

Equations (\ref{eq:layers}) for two-layer flow in the moving coordinate system have the form
\begin{equation}\label{eq:twolayers}
 \begin{array}{l}\displaystyle
    \rho_1u_{1t}+\rho_1(\beta u_1+\gamma U)u_{1x}+g\rho_1h_{1x}+g\rho_2h_{2x}
    +p_{0x}=-\mu_1(u_1+U), \\[3mm]\displaystyle
    \rho_2u_{2t}+\rho_2(\beta u_2+\gamma U)u_{2x}+p_{0x}
    =-\mu_2(u_2+U), \\[3mm]\displaystyle
    h_{1t}+(u_1 h_1)_x=0, \quad h_{2t}+(u_2 h_2)_x=0,
   \end{array}
 \end{equation}
where $\gamma=\beta-1$. The depth and velocity of the second layer are given by formulas
\[ h_2=H-h_1, \quad u_2=-(H-h_1)^{-1} u_1 h_1. \]
We linearize the momentum equations in (\ref{eq:twolayers}) and obtain the modified system
\[ \begin{array}{l}\displaystyle
    \rho_1u_{1t}+\rho_1\gamma U u_{1x}+g(\rho_1-\rho_2)h_{1x}+p_{0x}= -\mu_1(u_1+U), \\[2mm]\displaystyle
    \rho_2u_{2t}+\rho_2\gamma U u_{2x}+p_{0x}= -\mu_2(u_2+U).
   \end{array} \]
Using the change of variables mentioned above $s=\rho_1 u_1-\rho_2 u_2$ we obtain the motion equations in the evolutionary form
\begin{equation}\label{eq:evolution_lin_model}
   \begin{array}{l}\displaystyle
    s_t+\big(\gamma U s+g(\rho_1-\rho_2)h\big)_x=
    (\mu_2-\mu_1)U-\frac{\mu_2h+\mu_1(H-h)}{(H-h)\rho_1+h\rho_2}s\,,
    \\[3mm]\displaystyle
    h_t+(uh)_x=0,
   \end{array}
\end{equation}
where
\[ h=h_1, \quad u=u_1=\frac{(H-h)s}{(H-h)\rho_1+h\rho_2}\,. \]
It is not difficult to rewrite equations (\ref{eq:evolution_lin_model}) in the form $\bU_t+\bA \bU_x=\bF$, where $\bU=(s,h)^{\rm T}$ is the vector of unknown variables. The right part $\bF$ and matrix~$\bA$ have the following form
\[ \begin{array}{l}\displaystyle
    \bF=\Big(-\frac{(\mu_2h+\mu_1(H-h))s}{(H-h)\rho_1+h\rho_2}+
    (\mu_2-\mu_1)U,\quad 0\Big)^{\rm T}, \\[4mm]\displaystyle
   \bA=\left(\begin{array}{c c}
        \gamma U & g(\rho_1-\rho_2) \\
        \psi_s & \psi_h \\
    \end{array}
    \right ),\qquad  \psi=\frac{(H-h)hs}{(H-h)\rho_1+h\rho_2}.
    \end{array} \]
Direct calculations show that the matrix ${\bf A}$ has the eigenvalues
\[ \lambda_{1,2}=\frac{1}{2}\bigg(\gamma U+\frac{\partial \psi}{\partial h}\pm
   \sqrt{\Big(\gamma U-\frac{\partial \psi}{\partial h}\Big)^2+4g\big(\rho_1-\rho_2\big)
   \frac{\partial \psi}{\partial s}}\,\bigg). \]
Correspondingly the characteristics of system (\ref{eq:evolution_lin_model}) can be obtained from the solution of equation $dx/dt=\lambda_{1,2}$. Due to $0\leq h\leq H$ and $\psi'_s=\psi/s>~0$ system (\ref{eq:evolution_lin_model}) is hyperbolic in the case of $\rho_1\geq\rho_2$. Standard methods developed for integrating the hyperbolic conservation laws can be used for the numerical solution of equations (\ref{eq:evolution_lin_model}). We apply the simple and robust Nessyahu--Tadmor scheme for calculations of the interface position between two fluids with different physical properties.

If we assume that the summands containing the velocity derivatives are equal to zero in
 equations (\ref{eq:twolayers}) then we have Darcy-like model
\[ \begin{array}{l}\displaystyle
     g(\rho_1-\rho_2)h_{1x}+p_{0x}=-\mu_1(u_1+U), \quad p_{0x}=-\mu_2(u_2+U), \\[3mm]\displaystyle
     h_{1t}+(u_1 h_1)_x=0, \quad u_2=-(H-h_1)^{-1} u_1 h_1.
   \end{array} \]
We can eliminate the pressure from these equations and specify the velocity $u_1$ as the function of layer height $h=h_1$ and its derivative $h_x$. Now the system reduces to Burgers-like equation
\begin{equation}\label{eq:Darcy-like_model}
 h_t+\big(\Phi(h,h_x)\big)_x=0,
\end{equation}
where
\[ \Phi(h,h_x)=uh=
   \frac{(H-h)h}{(H-h)\mu_1+h\mu_2} \Big((\mu_2-\mu_1)U-g(\rho_1-\rho_2)h_x\Big).\]

If a fluid moves without gravity action ($g=0$) or without stratification ($\rho_1=\rho_2$) the equation obtained is replaced by the kinematic-wave model where $\Phi$ does not depend on $h_x$.

\begin{figure}[h!]
\begin{center}
\resizebox{0.96\textwidth}{!}{\includegraphics{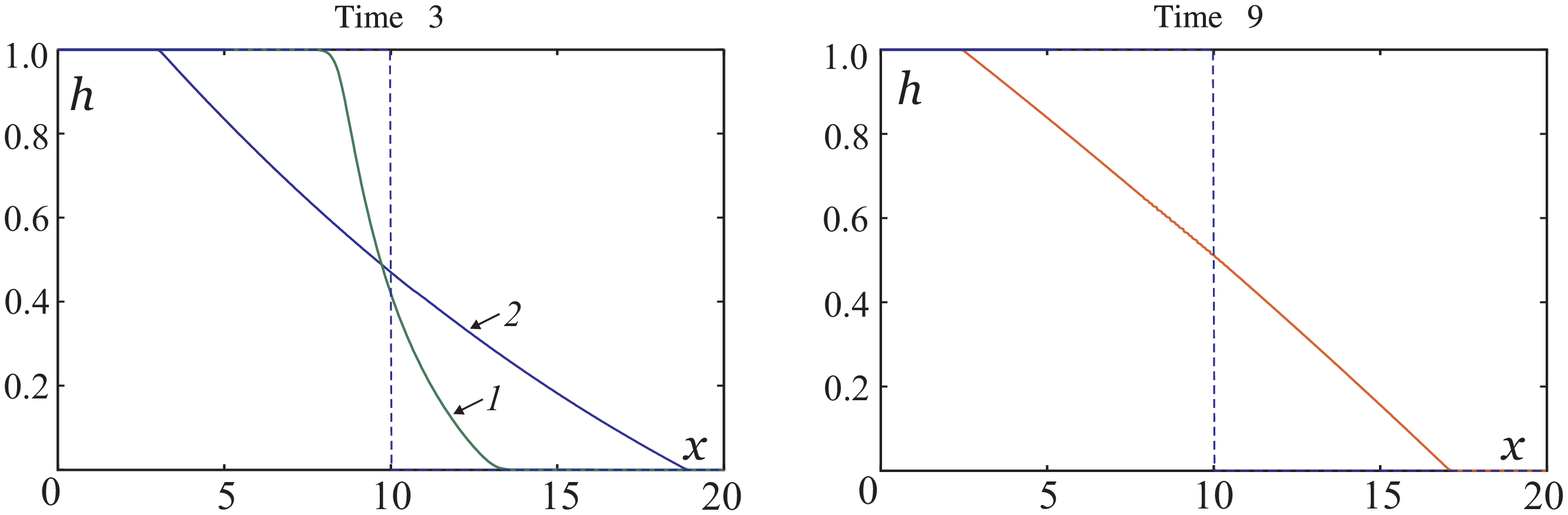}}\\[0pt]
\parbox{0.48\textwidth}{\caption{{\small The interface: $\mu_1=1$, $\rho_1=1$, $\mu_2=2$, $\rho_2=0.98$. Curve {\it 1} --- $g=0$, {\it 2} --- $g=980$}} \label{fig:4}}
\hfill
\parbox{0.48\textwidth}{\caption{{\small The interface: $\mu_1=2$, $\rho_1=1$, $\mu_2=1$, $\rho_2=0.98$, $g=980$}} \label{fig:5}}
\end{center}
\end{figure}
We use parameters $U=1$, $H=1$ and $\beta=6/5$ for calculations with respect to models (\ref{eq:evolution_lin_model}) and (\ref{eq:Darcy-like_model}). We assume that initial interface
position is as follows
\[ h\big|_{t=0} =
   \left\{
   \begin{array}{ll}
     1, & \quad x<x_0 \\[2mm]
     0, & \quad x>x_0.
   \end{array}
  \right. \]
It is shown in Figs.~\ref{fig:4} and~\ref{fig:5} by the dashed line. We suppose additionally $s(0,x)=0$ for equation (\ref{eq:evolution_lin_model}). Results for model (\ref{eq:Darcy-like_model}) are presented in Fig.~\ref{fig:4} at $t=3$. It corresponds to displacing more viscous fluid $(\mu_2=2)$ by less viscous one $(\mu_1=1)$. The flow without gravitation ($g=0$) is shown by curve~{\it 1}. In this case we have the Saffman--Taylor instability, i.e. the interface position changes due to less viscosity of displacing fluid. Note that in the case of displacing more viscous fluid $\mu_1>\mu_2$ the interface holds the initial position for all time at $g=0$. Curve~{\it 2} corresponds to $g=980$ and shows that positive stratification fades up the viscous finger velocity. Fig.~\ref{fig:5} displays Rayleigh--Taylor instability (more viscous and denser fluid displaces less viscous and less dense one due to gravitation) in the framework of model~(\ref{eq:Darcy-like_model}). Although displacing fluid is more viscous the interface deviates from the initial position because the gravity prevails for the parameters given.
\begin{figure}[h!]
\begin{center}
\resizebox{0.96\textwidth}{!}{\includegraphics{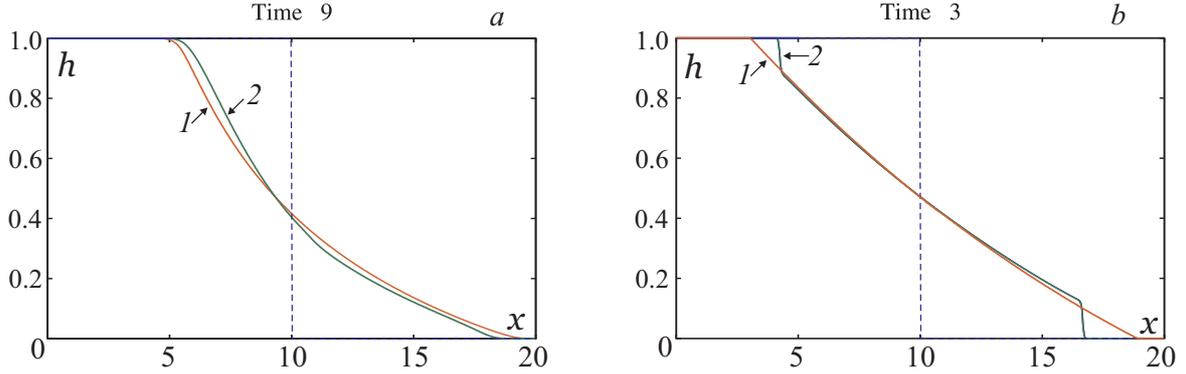}}\\[0pt]
{\caption{\small{The interface at $\mu_1=1$, $\rho_1=1$, $\mu_2=2$,
$\rho_2=0.98$; curve {\it 1} corresponds to
(\ref{eq:Darcy-like_model}), curve {\it 2} corresponds to
(\ref{eq:evolution_lin_model}); {\it a}
--- $g=0$, {\it b} --- $g=980$} } \label{fig:6}}
\end{center}
\end{figure}

Fig.~\ref{fig:6} presents comparison of calculations of the interface position with respect to models (\ref{eq:evolution_lin_model}) (curves {\it 2}) and (\ref{eq:Darcy-like_model}) (curves {\it 1}). Fig.~\ref{fig:6},~{\it a} corresponds to $t=9$ and demonstrates change of the interface
position without gravity action. It is not difficult to see that qualitative behavior of curves calculated with respect to different models coincides. If we change displacing and displaced fluids then instability does not develop and the interface does not move.

There is the interface position for the gravity action in Fig.~\ref{fig:6},~{\it b} at $t=3$. The calculation results coincide in the main flow region and differ in the vicinity of walls only. Note that flow structure for hyperbolic model (\ref{eq:evolution_lin_model}) has strong discontinuities adjacent to non-constant solution region. It is caused by the characteristics of equations (\ref{eq:evolution_lin_model}) are linearly degenerate at the initial data considered \cite{RYa78}. Truly the eigenvectors corresponding to the eigenvalues $\lambda_j$ ($j=1,2$) of matrix $\bA$ have the form $\br_j=(\lambda_j-\psi_h,\psi_s)^{\rm T}$. We construct the vectors $\bq_j=\nabla \lambda_j$. The gradient is calculated with respect to dependent variables $s$ and $h$. Direct calculations show that scalar products $\bq_j\cdot\br_j$ are equal to zero at $s=0$ and $h=1$ ($h=0$). It explains the qualitative behavior of system (\ref{eq:evolution_lin_model}) solution in Fig.~\ref{fig:6},~{\it
b}.

\section{Conclusion}

The class of shear flow stability is studied for model (\ref{eq:model}). It describes Hele-Shaw flows of an inhomogeneous incompressible fluid taking into account dependence of viscosity and
density on concentration. The latter holds along the trajectories. As a result of linearized equation solution the eigenvalues problem is formulated. In particular case this problem reduces to well-known Rayleigh equation. Analysis of two-layer flows of binary mixture with piecewise constant velocity, viscosity and density is provided (Fig.~\ref{fig:1}). Boundary conditions at the interface
(\ref{eq:kin-bc}) and (\ref{eq:din-bc}) are obtained. These formulas with equations (\ref{eq:psi-main}) lead to dispersion relation (\ref{eq:disp-rel}). Unlike a number of other papers the factor $\beta=6/5$ in front of convective terms is taken into account. This factor is generated at averaging through a cell gap. The dispersion curves are presented in Fig.~\ref{fig:2} and demonstrate that viscosity stabilizes the long-wave perturbations. If the factor $\beta\neq 1$
then flow with slip line (contact discontinuity) without viscosity effect can be stable in a certain range of parameters. The stability region extends for the stratified flows (Fig.~\ref{fig:3}). The growing perturbations exist for the wave numbers $k>k_*$, where $k_*$ is the critical number. Walls do not influence on qualitative character of results in the framework of the scheme considered.

With the help of equations of layered flows it is shown that the simplified models are very suitable for the interface position modeling. The calculations with respect to these models demonstrate that the interface is not stable when more viscous fluid is displaced by less viscous one (at least for fluid without density stratification). The hyperbolicity of linearized model (\ref{eq:evolution_lin_model}) is proved for two-layer fluid at certain density relation. In the case of model based on Darcy law (\ref{eq:Darcy-like_model}) it is shown that it can be reduced to one Burgers-like equation. The kinematic-wave model (Hopf equation) can be obtained for homogeneous fluid with respect to density or without gravity action flow. The numerical calculations of the interface position using models (\ref{eq:evolution_lin_model}) and (\ref{eq:Darcy-like_model}) display qualitative coincidence of the results. They differ in slower interface velocity close walls in the case of hyperbolic system (\ref{eq:evolution_lin_model}) only.

\section*{Acknowledgments}
The work is supported by Russian Foundation of Basic Research (project 14-31-50572), the Program of Leading Scientific Schools Supporting (project 2133.2014.1) and Integrating project of SB RAS
(44).

\end{document}